\documentstyle[preprint,aps,psfig]{revtex}
\tighten
\begin{document}
\title{Inhomogeneous nucleation in quark-hadron
phase transition}
\author{P. Shukla, A.K. Mohanty, and S.K. Gupta}
\address{ Nuclear Physics Division,
Bhabha Atomic Research Centre,\\
Trombay, Mumbai 400 085, India}
\author{Marcelo Gleiser}
\address{Department of Physics and Astronomy,
Dartmouth College,\\
Hanover, NH 03755, USA}
\maketitle
\footnote{To appear in Phys. Rev. C}
\begin{abstract}
\noindent
The effect of subcritical hadron bubbles
on a first-order quark-hadron phase transition is studied.
These subcritical hadron bubbles are created due to
thermal fluctuations, and can introduce a finite amount of
phase mixing (quark phase mixed with hadron phase)
even at and above the critical temperature.
For reasonable choices of surface tension and
correlation length, as obtained from the lattice QCD calculations,
we show that the amount of phase mixing at the critical
temperature remains below the percolation threshold.
Thus, as the system cools below the critical temperature,
the transition proceeds through the nucleation of critical-size
hadron bubbles from a metastable quark-gluon phase (QGP),
within an inhomogeneous background populated by an equilibrium
distribution of subcritical hadron bubbles.
The inhomogeneity of the medium
results in a substantial reduction of the nucleation barrier
for critical bubbles.
Using the corrected nucleation barrier, we estimate the amount of
supercooling for different parameters controlling the phase
transition, and briefly discuss its implications to cosmology and
heavy-ion collisions.
\\
PACS number(s): 12.38.Mh, 64.60.Qb, 05.70.Fh, 25.75-q, 98.80.Cq
\end{abstract}

\newpage
\section {Introduction}
The hadronization of Quark Gluon Plasma (QGP) possibly
produced in the early universe or expected to be formed
in relativistic heavy-ion collisions has been the focus of
much attention during the past few years.
However, the mechanism of hadronization (QCD phase transition)
remains an open question.
The prediction of lattice QCD on the order
of the transition is still unclear, if physical masses
for quarks are used \cite{LQCD1}. Quenched
QCD (no dynamical quarks) shows a first-order phase transition, albeit
a weak one, with small surface tension and latent heat \cite{LQCD2}.
Assuming the transition to be first-order,
homogeneous nucleation theory \cite{LANGER67,LANGER73}
has been invoked extensively to
study the dynamics of the  quark hadron phase transition
both in the context of early universe as well as for the plasma
produced during relativistic heavy-ion collisions
\cite{CSER1,CSER2,ZABRO,ZABRO1,SHUK,FULLER,MEYER,IGNAT1}.
In this picture, the transition is initiated by the nucleation
of critical-size hadron bubbles from a supercooled metastable QGP phase.
These hadron bubbles can grow against surface tension, converting
the QGP phase into the hadron phase as the
temperature drops below the critical temperature,
$T_C$. This is indeed the case for a sufficiently
strong first order transition, where the assumption of
a homogeneous background of QGP is justified
at the time when the nucleation begins.
However, for a weak enough transition, the QGP phase
may not remain in a pure homogeneous state even at $T=T_C$, due
to pre-transitional phenomena.
For temperatures much above $T_C$, matter is in a pure
QGP phase with the effective potential exhibiting one minimum at
$\phi$=0. Here $\phi$ is an effective scalar order parameter generally used to
model the effective potential describing the dynamics of a phase
transition. As the plasma expands and
cools to some temperature $T_1$, an inflection
point is developed away from the origin which on further cooling
separates into a maximum at $\phi$=$\phi_m$
and a local minimum at $\phi=\phi_h$, corresponding to the hadron phase.
At $T=T_C$, the potential is
degenerate with a barrier separating the two phases.
``Pre-transitional phenomena'' refers to nonperturbative dynamical
effects above $T_C$
in the range $T_C\le T \le T_1$. Such phenomena are known to occur in
several areas of condensed matter physics, as in the case of isotropic to
nematic phase transition in liquid crystals\cite{COND}, and are
also expected in the cosmological electroweak phase
transition leading to large phase mixing at $T=T_C$\cite{GK}. In such
cases, the phase transition may proceed either through
percolation \cite{GK,PERCO}
or, if the phase mixing is below the percolation threshold, by the
nucleation of critical bubbles in the background of isolated
hadronic domains, which
grow as $T$ drops below $T_C$. In either case, the kinetics is quite different
from what is expected on
the basis of homogeneous nucleation \cite{GH}. We will argue that, for a
wide range of physical parameters, a large amount of thermal phase mixing
at $T=T_C$ is expected to occur during the quark-hadron phase transition
in the early universe, as well for the plasma produced in heavy-ion
collisions \cite{AGAR}. For high enough temperatures and low enough
cooling, large-amplitude thermal fluctuations will populate
the new minimum at $\phi=\phi_h$ in the range $T_C \le T \le T_1$.
Although these fluctuations which are in the form of
subcritical hadron bubbles will
shrink and finally disappear, there will always be some non-zero
number density of hadron bubbles at a given temperature $T$.
In this work, we study the equilibrium density distribution of subcritical
hadron bubbles for a wide spectrum of very weak to very strong first order
QCD phase transition,
using the formalism developed in Refs. \cite{GKW,GELMINI,GHK}.
It is found that the density of subcritical hadron bubbles builds up
faster as the transition becomes weaker, leading, in some cases,
to complete phase mixing
at $T=T_C$. Further, using reasonable values for the surface tension and
correlation length as obtained from lattice QCD calculations, we find that
(although large)
the amount of phase mixing remains below the percolation
threshold. Therefore, the quark-hadron
phase transition will begin with the nucleation of critical-size
hadron bubbles from a supercooled and inhomogeneous background of quark-gluon
plasma. Since the background contains subcritical hadron bubbles,
the homogeneous theory of nucleation needs to be modified. In Ref. \cite{GH},
an approximate method was suggested to incorporate this inhomogeneity  by
modelling subcritical bubbles as Gaussian fluctuations,
resulting in a large reduction in the nucleation barrier. Here, we will
study inhomogeneous nucleation in the framework
of homogeneous theory, but  with a reduced nucleation barrier that accounts
for the inhomogeneity of the medium.
Finally, we also briefly discuss possible implications of inhomogeneous
nucleation to relativistic heavy-ion collisions and cosmology.

The paper is organized as follows. In the next section, we begin with the
discussion of a quartic double-well potential
used to describe the
dynamics of a first-order quark-hadron
phase transition. The parameters of the potential
are obtained in terms of relevant physical quantities such as critical
temperature, surface tension and correlation length.
 In section III, we estimate the equilibrium
fraction of subcritical hadron bubbles from
very weak to strong first-order phase transitions.
We also estimate the reduction in the nucleation barrier by incorporating
the presence
of subcritical bubbles in the medium. Using this reduced barrier, we study
nucleation and supercooling in section IV. Finally, we present our
conclusions in section V.

\section{Parameterization of the effective potential}
We consider a general form of the potential (or equivalently, the
homogeneous part of the Helmholtz free energy density) to study the
quark-hadron phase transition in terms of a real scalar order parameter
$\phi$ given by

\begin{eqnarray}\label{vt}
V(\phi, T) &=& a(T)\,\phi^2 - b \,T \,\phi^{3 } +  c \, \phi^4,
\end{eqnarray}
where $b$ and $c$ are  positive constants.
Ignatius et. al. \cite{IGNAT2} use this parameterization to
describe the phase transition from
a QGP (symmetric phase) to a hadron phase (broken symmetry).
The meaning of $\phi$ is obvious for a symmetry-breaking transition,
but the same description can be used if no symmetry is involved.
The order parameter could then be related to energy or entropy density.
The parameters $a$, $b$ and $c$ are determined in terms of surface tension $(\sigma)$,
correlation length $(\xi)$ and critical temperature $(T_C)$.
The potential has two minima, one at $\phi_q = 0$ and the other
at $\phi_h= (3 b T + \sqrt{9 b^2 T^2 -32 a c})/8c$,
which in our case will represent quark and hadron phases respectively.
These phases are separated by a maximum defined by
$\phi_m= (3 b T - \sqrt{9 b^2 T^2 -32 a c})/8c$.
At $T=T_C$,
\begin{eqnarray}\label{deg}
V(\phi_q, T_C) = V(\phi_h, T_C) = 0,
\end{eqnarray}
having the required degeneracy. The above condition yields,

\begin{eqnarray}\label{relat}
a(T_C)=b^2 T_C^2/4 c, \hspace{.1in}
\phi_h(T_C)=b T_C/2 c
\hspace{.1in}{\rm and} \hspace{.1in}
\phi_m(T_C)=b T_C/4 c.
\end{eqnarray}
Using these relations, the barrier height at $T_C$ can be
written as
\begin{eqnarray}\label{hb}
V_b = b^4 T_C^4/ 256 c^3.
\end{eqnarray}
Therefore, if the parameter $c$ is kept
fixed, $b$ can be varied to characterize a wide spectrum of very
weak to very strong first-order phase transitions.
The transition is strong enough for
large $V_b$ and very weak or close to second order as
$V_b \rightarrow 0$.
In the following, we relate the
parameters $b$ and $c$ to the surface tension and the
correlation length in the quark phase.
The surface tension can be defined as the one dimensional action
given by,

\begin{eqnarray}\label{}
\sigma = \int dx \left[ \frac{1}{2}
   \left( \frac{\partial \phi}{\partial x}\right)^2 + V(\phi) \right].
\end{eqnarray}
Under the thin-wall limit, $\Delta = |V(0) - V(\phi_h)| \rightarrow 0$,
the surface tension
can be expressed as \cite{LINDE}

\begin{eqnarray}\label{sig}
\sigma &=& \int_0^{\phi_h} d\phi \sqrt{2 V(\phi)}, \nonumber \\
         &=& \frac{\sqrt{2}}{48} \frac{b^3 T_C^3}{c^{5/2}}.
\end{eqnarray}
Similarly, the correlation length around the quark phase is
obtained using $\xi_q =  1/\sqrt{V''(\phi)} |_{\phi=0}$
$=1/\sqrt{2 a(T)}$. At the critical temperature,
using Eq.~(\ref{relat}), we get

\begin{eqnarray}\label{xi}
\xi_q(T_C) = \frac{\sqrt{2 c}}{b T_C}.
\end{eqnarray}
{}From Eqs.~(\ref{sig}) and (\ref{xi}) we get

\begin{eqnarray}\label{cb}
c =  \frac{1}{12\xi_q^3 \sigma}, \hspace{.2in}
b^2 = \frac{1}{6\xi_q^5 \sigma T_C^2},
\end{eqnarray}
in terms of the values of $\sigma$ and $\xi_q$ at $T_C$.
The barrier height $V_b$ can now be written as

\begin{eqnarray}\label{hbs}
V_b = {3\over 16} {\sigma \over \xi_q(T_C)}.
\end{eqnarray}
Thus, the barrier height is proportional to the ratio
$\sigma/\xi_q$. The transition becomes very weak as $\sigma$ decreases
and $\xi_q$ increases. Here, we fix $\xi_q=0.5$ fm at $T=T_C$ and
vary $\sigma$ to investigate phase transitions with different strengths.
The temperature dependence of $a$ is deduced by equating
the depth of the second minimum with the
the pressure difference $\Delta P$ between the two phases at
all temperatures. This yields an equation

\begin{eqnarray}\label{atemp}
\Delta P &=& p_h-p_q  \nonumber \\
         &=& V(0)-V(\phi_h)  \nonumber \\
         &=&- \left( a(T) - b T \phi_h + c \phi_h^2\right)\phi_h^2
\end{eqnarray}
which is solved to get the parameter $a(T)$, giving
the temperature dependence of $\xi_q$. The surface tension
will also have small temperature dependence which we ignore,
as we are not going too far from the critical temperature.
Thus, we have parameterized the free-energy density in terms of the
surface tension, correlation length, critical temperature
and equation of state, which can be obtained from lattice QCD calculations.
The bag equation of state which is a good depiction of
the lattice results is used to calculate
the quark/hadron pressure $p_{q/h}$ as follows

\begin{eqnarray}\label{}
p_q=a_q \,T^4-B, \hspace{.2in} p_h=a_h \,T^4,
\end{eqnarray}
where $B=(a_q-a_h)\, T_C^4$ is the bag constant.
The quark phase is assumed to consist of a massless gas of $u$ and $d$ quarks
and gluons, while the hadron phase contains massless pions. Thus, the
coefficients $a_q$ and $a_h$ are given by
$a_q = 37 \pi^2/90$ and $a_h = 3 \pi^2/90$. The critical
temperature is taken as $T_C=160$ MeV.

Fig. 1 shows the plot of $V(\phi)$ as a function of $\phi$ at three different
temperatures for a typical value of
$\sigma$ = 30 MeV/fm$^2$ and $\xi_q(T_C) = 0.5$ fm.
At $T=T_C$, the potential is degenerate with a large barrier that separates
the two phases. Below $T_C$, the phase $\phi=\phi_h$ has lower free-energy
density, and the QGP phase becomes metastable.
Above $T_C$, the potential has a metastable minima at $\phi=\phi_h$
(hadron phase) as long as $T$ remains below $T_1$.
The temperature $T_1$ [at which $\phi_h=\phi_m$ and
$9 b^2 T_1^2 = 32 a(T_1)\,c$] can be obtained analytically by
solving Eq.~(\ref{atemp}) as,
\begin{eqnarray}
T_1 = \left[\frac{B}{B - {27\over 16} V_b} \right]^{1/4} \, T_C.
\end{eqnarray}
It may be mentioned here that the dynamics of the phase transition has
also been studied in Ref.~\cite{AGAR} using
a different form of the potential which has been parameterized as a fourth
order polynomial in the energy density \cite{CSER1}.
This form is unsuitable over a wide range of temperatures due to
the persistence of metastability at much above and below $T_C$.

\section{Model for large-amplitude fluctuations}

We closely follow the work of Refs.
\cite{GH,GELMINI,GHK} to estimate the equilibrium
density distribution of subcritical hadron bubbles by
modeling them as Gaussian fluctuations
with amplitude $\phi_A$ and radius $R$

\begin{eqnarray}\label{gau}
\phi_{q \rightarrow h}(r) = \phi_A e^{-r^2/R^2} \hspace{.1in} {\rm and}
\hspace{.1in}
\phi_{h \rightarrow q}(r) = \phi_A \left( 1 - e^{-r^2/R^2}\right).
\end{eqnarray}
The amplitude $\phi_A$ is the value of the field at the bubble's core away
from the quark phase.
For smooth interpolation between the two phases in the system,
$\phi_A \ge \phi_m$.
The free energy of a given configuration can then be found by using
the general formula \cite{LINDE},

\begin{eqnarray}\label{fs}
F = \int d^3r \left[ \frac{1}{2}(\nabla \phi(r))^2 +
      V\left(\phi(r)\right)\right].
\end{eqnarray}
Using Eq.~(\ref{gau}) and  Eq.~(\ref{vt}) in Eq.~(\ref{fs}) we get

\begin{eqnarray}\label{}
F_{q \rightarrow h} \equiv F_h
= \alpha_h \, R + \beta_h \, R^3 \hspace{.1in} {\rm and}
\hspace{.1in}
F_{h \rightarrow q} \equiv F_q
= \alpha_q \, R + \beta_q \, R^3,
\end{eqnarray}
where $\alpha_h$, $\beta_h$, $\alpha_q$ and $\beta_q$ are given by

\begin{eqnarray}\label{}
\alpha_h = \alpha_q=\frac{3\sqrt{2}}{8} \pi^{3/2} \phi_A^2, \hspace{.2in}
\beta_h =\left[ \frac{\sqrt{2} a}{4}  - \frac{\sqrt{3} b T}{9} \phi_A
         +  \frac{c}{8} \phi_A^2  \right] \pi^{3/2} \phi_A^2
\end{eqnarray}
and
\begin{eqnarray}\label{}
\beta_q &=& \left({\sqrt{2}\over 4}-2\right) a   \pi^{3/2} \phi_A^2
     - \left(-{\sqrt{3}\over 9}-3 + {3\sqrt{2}\over4}\right)
             b T  \pi^{3/2} \phi_A^3 \nonumber \\
     &+&  \left( {1\over 8}+{3\sqrt{2}\over 2}-{4\sqrt{3}\over 9}
      -4\right) c \pi^{3/2} \phi_A^4.
\end{eqnarray}

It may be mentioned here that $\alpha_h(=\alpha_q)$ is positive and is much
greater than $\beta_{h(q)}$. Therefore, the free energy grows linearly
for small values of $R$. Further, hadron bubbles of all configurations
will be subcritical as long as $\beta_{h(q)}$ is positive. At $T=T_C$,
both $\beta_h$ and $\beta_q$
are positive for all amplitudes. However, below $T_C$, $\beta_h$ may become
negative for some values of $\phi_A$. For such configurations, the free
energy has a maximum at $R_m=\sqrt{\alpha_h/3\beta_h}$ and these bubbles
are not strictly subcritical. The same is true for $\beta_q$ above $T_C$.
We thus restrict the amplitudes $\phi_A$ to the range where
$\beta_{h(q)}$ is positive. If not exactly the same, the limits of
integration $\phi_{min}$ and $\phi_{max}$ for $\phi_A$
are found to be quite close to $\phi_m$ and $\phi_h$ respectively.

\subsection {Equilibrium fraction of subcritical bubbles}

There will be fluctuations from quark to hadron phase and back.
To obtain the number density
$n_A$ of subcritical bubbles, we define the distribution function
$f\equiv \partial^{2}n_A/\partial R\partial \phi_A$ where
$f(R,\phi_A,t)dR d\phi_A$ is the number density of bubbles
with radius between $R$ and $R+dR$ and amplitude between
$\phi_A$ and $\phi_A + d\phi_A$ at time $t$. It satisfies the
Boltzmann equation \cite{GH,GHK}

\begin{eqnarray}\label{boltz}
\frac{\partial f(R,\phi_A,t)}{\partial t} =
- |v|\frac{\partial f}{\partial R} + (1-\gamma )G_{h}
 - \gamma G_{q}.
\end{eqnarray}
The first term on the RHS is the shrinking term. Here,
$|v|$ is the shrinking velocity, which we assume to be given
by the velocity of sound ($=1/\sqrt{3}$) in a massless gas.
The second term is the
nucleation term where $G$ is the Gibbs distribution function defined
as $\Gamma=\int dRd\phi G$. Here $\Gamma_{h}$ is the nucleation
rate per unit volume of subcritical bubbles from the quark phase to
the hadron phase.
Similarly $\Gamma_{q}$ is the corresponding rate from the hadron
phase to the quark phase.
The factor $\gamma$ is defined
as the fraction of volume in the hadron phase and is obtained by summing over
subcritical bubbles of all amplitudes and radii within this phase. The Gibbs
distribution function is defined as
\cite{GELMINI,LINDE}

\begin{eqnarray}\label{}
G_{h/q} = A\, T^4 \, e^{-F_{h/q} (R,\phi_A)/T},
\end{eqnarray}
where $A$ is of
${\cal {O} }\sim 1$ \cite{LINDE}.

If the equilibration time scale is smaller than the expansion
time scale of the system, we can obtain
the equilibrium number density of subcritical bubbles
by solving Eq.~(\ref{boltz}) with $\partial f/\partial t =0$.
Since the early universe expands at a much slower rate \cite{FULLER,MEYER},
the above assumption is quite reasonable in the context of the cosmological
QCD phase transition. However, QGP produced during heavy ion collision
may expand at a faster rate as compared to the early universe.
In this case, it is possible that the density distribution of the
subcritical bubbles will not attain full equilibrium.
For simplicity, we will assume an equilibrium situation so that
the present results on the fraction of subcritical bubbles
and phase mixing can be considered as an upper limit.
Using the boundary condition $f(R\rightarrow \infty)=0$, we get the
equilibrium distribution given by

\begin{eqnarray}
f(R, \phi_A, T) = (1-\gamma) \, W_S(R,\phi_A, T) -
              \gamma \, W_T(R,\phi_A, T),
\end{eqnarray}
where
\begin{eqnarray}
W_S(R,\phi_A, T)&=& (A/|v|) T^4
         \int_R^{\infty} e^{-(\alpha_h R' + \beta_h R'^3)/T} dR',
                        \nonumber \\
W_T(R,\phi_A, T)&=& (A/|v|) T^4
         \int_R^{\infty} e^{-(\alpha_q R' + \beta_q R'^3)/T} dR'.
\end{eqnarray}
The equilibrium fraction $\gamma$ of volume occupied by
subcritical bubbles is given by,

\begin{eqnarray}
\gamma(\phi_{min},\phi_{max},R_{\rm min},R_{\rm max}) &=&
      \int_{\phi_{min}}^{\phi_{max}} \int_{R_{min}}^{R_{\rm max}}
       \frac{4\pi}{3} R^3  f(R, \phi_A, T) dR d\phi_A,
\end{eqnarray}
which is solved to get
\begin{equation}
\gamma =\frac{I_S}{1+I_S+I_T},
\end{equation}
where
\begin{equation}
I_{S(T)}= \int_{\phi_{min}}^{\phi_{max}}
     \int_{R_{\rm min}}^{R_{\rm max}}
   \frac{4\pi}{3}  R^3 W_{S(T)} (R, \phi_A, T) dR d\phi_A~.
\end{equation}

Here, $\phi_{min}$ and $\phi_{max}$ define the range within
which both $\beta_h$
and $\beta_q$ are positive. $R_{min}$ is the smallest radius of
the subcritical
bubbles taken as $\xi_q$, the correlation length of the fluctuations.
The $R$ integration should be carried out over all bubbles with radii
from $R_{min}=\xi_q$ to $R_{max}=\infty$.
For very weak transitions, both $\alpha$ and $\beta$ are very small
and the $R$ integration may not have good convergence. However, we found
that the $\gamma$ value is maximized when $R_{max}$ is about 3 to 4 fm.
Therefore, we use $R_{max}$=3.5 fm. This is a reasonable choice as
bubbles with $R \sim \xi_q$
will be statistically dominant
and larger fluctuations have larger free
energy and are exponentially suppressed.

Fig.~2 shows the plot of the subcritical hadron fraction
$\gamma$ as a function of $\sigma$ at $T=T_C$
and at a fixed value of $\xi_q(T_C)=0.5$ fm.
The fraction $\gamma$ has been estimated (dashed curve) assuming
that, for a degenerate potential, $G_h \simeq G_q$, as in
Ref.~\cite{GH}.
This assumption is valid only for the configuration for which $\phi_A=\phi_h$.
However, when we include
other configurations in the range $\phi_{min}$ to $\phi_{max}$, the integral
$I_T$ turns out to be always higher than $I_S$ at $T_C$. Therefore, $\gamma$
obtained using $G_h\neq G_q$ is always lower than
when the approximation $G_h=G_q$ is used. In both cases,
the value of $\gamma$ increases with decreasing $\sigma$ i.e.
when the transition becomes weak.
It may be mentioned here that
as per lattice QCD calculations without dynamical quarks
\cite{LQCD2}, $\sigma$ lies between 2 MeV/fm$^2$  and 10 MeV/fm$^2$.
There would be $15\%$ to $30\%$ phase mixing
corresponding to these $\sigma$ values, which is still
below
the percolation threshold ($\gamma \le 0.3$).
If $\gamma > 0.3$, the two phases will mix completely, the
mean-field approximation for the potential breaks down,
and the phase transition may proceed through percolation \cite{PERCO,GHK}.
However, for a surface tension in the range 2 MeV/fm$^2\le \sigma \le$
10 MeV/fm$^2$,
the phase transition will proceed through
the formation of critical-size hadron bubbles from a supercooled metastable
QGP phase. Since the QGP phase is no longer homogeneous, the dynamics
of the phase transition will be quite different from what is expected on the
basis of homogeneous nucleation theory \cite{GH}.
We refer to it as ``inhomogeneous nucleation.''

We would also like to mention here that the present results are in
disagreement with the findings of Ref.~\cite{AGAR}, where a large
fraction of subcritical hadron phase was found at and above $T_C$.
This scenario is highly unrealistic and probably could
be due to the choice of the potential parameterization, which shows a
metastable hadron phase much above $T_C$. Therefore, the authors of
Ref. \cite{AGAR}
found a finite fraction of hadron phase at temperatures
as high as twice $T_C$.
Furthermore, the value of $\gamma$ strongly depends on how
the shrinking term is incorporated in the calculation.
In our case, it is proportional to the
gradient $(\partial f/\partial R)$ that appears in the
kinetic equation~(\ref{boltz}) in a natural way, whereas in Ref. \cite{AGAR},
a specific assumption is made to
take into account the shrinking of the hadronic volume.

\subsection {The total free energy of subcritical bubbles and the
            nucleation barrier}

The nucleation rate in the standard theory
\cite{LANGER67,LANGER73} which neglects phase mixing,
is given by

\begin{eqnarray}\label{rate}
I \simeq A T^4 \, e^{-F_{\rm C}/T}.
\end{eqnarray}
Here $F_{\rm C}$ is free energy needed to form
a critical bubble in the homogeneous metastable
background. For an arbitrary thin-wall spherical
bubble of radius $R$ and amplitude $\phi_{\rm thin}{\
\lower-1.2pt\vbox{\hbox{\rlap{$<$}\lower5pt\vbox{\hbox{$\sim$}}}}\ } \phi_h$,
the free energy of the bubble takes the well-known form

\begin{eqnarray}\label{fthin}
F_{\rm thin}(R) &=&  - \frac{4 \pi}{3}R^3 \Delta V +  4\pi R^2 \sigma .
\end{eqnarray}
In the above, $\Delta V$ is defined as the difference in
free-energy density between the background medium and the
bubble's interior. For a homogeneous background (metastable), we can write,

\begin{eqnarray}\label{}
\Delta V \equiv \Delta V_{0} = V(0) - V(\phi_h).
\end{eqnarray}
If there is significant phase mixing in the background metastable state,
its free energy is no longer $V(0)$. One must also account for the free
energy density of the nonperturbative large amplitude fluctuations.
Following Ref.~\cite{GH}, we write the free energy density of the metastable
state as $V(0)+{\cal F}_{\rm sc}$, where ${\cal F}_{\rm sc}$ is the
extra free energy density which can be estimated
from the density distribution of subcritical bubbles as follows:

\begin{eqnarray}\label{calF}
{\cal F}_{\rm sc} &\approx&  \int_{\phi_{min}}^{\phi_{max}}
   \int_{R_{\rm min}}^{R_{\rm max}} F_h(R, \phi_A, T) \,\,
   f(R, \phi_A, T)  dR d\phi_A ,  \nonumber \\
 &=&  (1-\gamma)\,\int_{\phi_{min}}^{\phi_{max}}
   \int_{R_{\rm min}}^{R_{\rm max}}  F_h \, W_S \, dR d\phi_A
 -\gamma \,\int_{\phi_{min}}^{\phi_{max}}
   \int_{R_{\rm min}}^{R_{\rm max}}  F_h \, W_T \, dR d\phi_A.
\end{eqnarray}
Once we know the hadronic fraction $\gamma$ and the free
energy $F_h$ for a bubble of a given radius $R$ and
amplitude $\phi_A$, we can estimate the free-energy density correction due
to the presence of Gaussian subcritical bubbles.

Since, for a critical size bubble,
$\partial F/\partial R|_{R_{C}}=0$,
we can use Eq.~(\ref{fthin}) to obtain the free energy needed to
form a thin-wall critical bubble in a background of subcritical bubbles,

\begin{eqnarray}\label{barr}
F_{C} = \frac{4 \pi}{3} \sigma R_{C}^2, \hspace{.1in}
R_{C} = \frac{2 \sigma}{\Delta V_{0} + {\cal F}_{\rm sc}}.
\end{eqnarray}
For a very strong first-order
phase transition, the subcritical bubbles are suppressed
$({\cal F}_{\rm sc} \rightarrow 0)$, and both $F_C$ and $R_C$
approach the homogeneous background expression. However, in the
presence of subcritical bubbles, extra free energy becomes available in the
medium, reducing the nucleation barrier.
In other words, the extra background energy enhances the nucleation
of critical bubbles.
To illustrate this, we have plotted $F_C/T$ and $\gamma$ as a function
of $T/T_C$ in Figs.~3 to 5 with $\sigma$ values of
50 MeV/fm$^2$, 30 MeV/fm$^2$ and 10 MeV/fm$^2$,
respectively, which are widely used in the literature.
As evident, with decreasing temperature,
the nucleation barrier decreases and
the subcritical hadron fraction $\gamma$ increases.
The reduction in barrier height
due to ${\cal F}_{\rm sc}$ (or due to
$\gamma$ ) is more significant for lower
values of $\sigma$, corresponding to a weaker transition.
Since the height of the nucletaion barrier decreases,
the nucleation rate will also be enhanced,
reducing the amount of supercooling further.
The time evolution of the
temperature and the supercooling are discussed in the next
section.

\section{Nucleation and Supercooling}

As mentioned before, the background metastable state is
inhomogeneous due to subcritical hadron bubbles.
It is now possible to study the
kinetics of the nucleation of critical hadron bubbles
using the corrected nucleation rate, as obtained in the
previous section.
In the present work, the prefactor in the nucleation rate is
taken as $AT^4$ [see Eq.~(\ref{rate})].
In our previous work, \cite{SHUK}, we have used a prefactor
derived by Csernai and Kapusta \cite{CSER2} for a
dissipative QGP. In Ref.~\cite{RUGGERI}, Ruggeri and Friedman
had derived a prefactor for a non-dissipative QGP.
Recently, using a more general formalism,
we have also derived a prefactor
\cite{DYNA} which has both dissipative and non-dissipative
components corresponding to Ref.~\cite{CSER2} and
Ref.~\cite{RUGGERI}, respectively.
However, for consistency with the
subcritical formalism, we use a more generic
form $I_0=AT^4$, with $A$ a constant of order unity, as
used in many studies of quark-hadron phase transition
(see, for example, Refs. \cite{FULLER,MEYER}). The question of how to estimate
the prefactor appearing in the nucleation rate of subcritical bubbles
remais open.
Using the nucleation rate $I(T)$,
the fraction $h$ of space which has been converted to hadron phase
due to nucleation of critical bubbles and their growth
can be calculated.
If the system cools to $T_C$ at a proper time $\tau_c$, then at some
later time $\tau$ the fraction $h$ is given by \cite{CSER2},

\begin{eqnarray}\label{frac}
h(\tau) = \int_{\tau_c}^\tau d\tau' I\left(T(\tau')\right)
         \left[1 - h(\tau') \right] V(\tau',\tau).
\end{eqnarray}
Here, $V(\tau',\tau)$ is the volume of a critical bubble at
time $\tau$ which was nucleated at an earlier
time $\tau'$; this takes into account the bubble growth.
The factor $\left[1 - h(\tau)\right]$
accounts for the available space for new bubbles to nucleate.
The model for bubble growth is simply taken as \cite{WEIN}

\begin{eqnarray}\label{}
V(\tau',\tau) = \frac{4\pi}{3} \left( R_C(T(\tau')) +
\int_{\tau'}^\tau d\tau'' v(T(\tau'')) \right) ^3,
\end{eqnarray}
where $v(T)=3 [1 - T/T_c]^{3/2}$ is the velocity of the
bubble growth at temperature $T$ \cite{CSER2,MILLER}.
The evolution of the energy density in 1+1
dimensions is given by

\begin{eqnarray}\label{hydro}
\frac{d e}{d\tau} + \frac{\omega}{\tau} = 0.
\end{eqnarray}
 The energy density $e$, enthalpy density $\omega$ and the
pressure $p$ in pure QGP and hadron phases are
given by the bag model equation of state.
In the transition region, the  $e$ and $\omega$
at a time $\tau$ can be written in terms of the hadronic fraction as

\begin{eqnarray}\label{mixed}
e(\tau)  =   e_q(T) + [e_h(T)-e_q(T)] h(\tau),
            \nonumber \\
\omega(\tau)  =   \omega_q(T) + [\omega_h(T)-\omega_q(T)] h(\tau).
\end{eqnarray}
Equations~(\ref{frac}), (\ref{hydro}), and (\ref{mixed}) are
solved to get the temperature as a function of time
in the mixed phase \cite{SHUK} with the initial conditions for
temperature $T_0=250$ MeV and proper time
$\tau_0=1$ fm/c at $T_C$=160 MeV.
After getting $T$ and $h$ as a function of time,
the density of nucleating bubbles at a time $\tau$
can be obtained in our model as

\begin{eqnarray}\label{}
N(\tau) = \int_{\tau_c}^\tau d\tau' I\left(T(\tau')\right)
         \left[1 - h(\tau') \right].
\end{eqnarray}
The density $N$ would increase as the temperature drops below
$T_c$ and would ultimately saturate as $h$ increases.

 Figure~6 shows the temperature variation as a function of proper
time $\tau$ at $\sigma=50$ MeV/fm$^2$.  As the system
cools below $T_C$, the nucleation barrier decreases
and also $\gamma$ increases.
If only homogeneous nucleation (dashed curve) is considered, the system
will supercool up to 0.945 $T_C$. At this temperature, the hadronic fraction
$\gamma$ has reached 10 \% (See Fig.~3),
which corrects the amount of supercooling (solid curve)
by about $\sim$10 \% (up to 0.95 $T_C$).
Figure~7 shows a similar study at $\sigma$ = 30 MeV/fm$^2$.
Since the nucleation barrier reduces with
decreasing $\sigma$, the system supercools only up to 0.98 $T_C$ under homogeneous
nucleation. The hadronic fraction $\gamma$ corresponding to this value
is $\sim 12-13$ \%
(See Fig.~4) which reduces the amount of supercooling by
about $\sim 20 \%$ (up to 0.984 $T_C$).
For $\sigma$ around
10 MeV/fm$^2$, the supercooling will be reduced further
(up to $\sim 0.997 T_C$).
Lattice QCD calculations predict a surface tension
even smaller than 10 MeV/fm$^2$, indicating a very weak
first order transition. Although supercooling will be reduced further
with decreasing $\sigma$, we do not use very small $\sigma$ due to
increased numerical
inaccuracy.
Further, it may be mentioned here that, although
the fraction $\gamma$ grows
with decreasing $\sigma$, we never
encountered $\gamma$ greater than 0.3: we remained
within the sub-percolation regime throughout our analysis.

Apart from $\sigma$, the amount
of supercooling also depends on $\tau_c$, the time taken by the system
to cool from $T_0$ to $T_C$. In QGP
phase, the solution of Eq. (\ref{hydro}) $(T^3\tau$ =constant) predicts $\tau_c=\tau_0(T_0/T_C)^3$.
The choice of $\tau_0$ = 1fm, $T_0$=250 MeV and $T_C$=160 MeV results in $\tau_c$=3.8 fm/c.
However, formation of QGP with higher initial temperature (as high as 3 to 4 times
$T_C$ resulting in large $\tau_c$) can not be ruled out at RHIC and LHC energies
\cite{SHU}. Therefore, we have also studied the effect of $\tau_c$ on supercooling,
specifically, on the hadronization rate as well as on the density of nucleating
bubbles. Figs. 8(a) and 8(b) show the plots of $N(\tau)$ and $h(\tau)$ as a
function of $\tau$ for two typical values of $\tau_c$ (3.8 fm/c and 25 fm/c)
corresponding to  $\sigma$=10 MeV/$fm^2$ both with (solid curve) and without
(dashed curve) inhomogeneity corrections. A general observation (both with and
without correction) is that the amount of supercooling,
the rates of hadronization and bubble nucleation
are reduced when $\tau_c$ becomes larger. Although supercooling reduces
with increasing $\tau_c$,
the system will get reheated at an earlier temperature and also
will encounter a larger nucleation
barrier as compared to the case when $\tau_c$
is small.
As a result, the rate of hadronization
and also the rate of increase of density
of the nucleating
hadron bubbles will proceed at a slower rate  when $\tau_c$ is large.
However, the
reverse happens when the inhomogeneity correction is applied.
Even though the medium gets heated up earlier,
the reduction in the barrier height
is quite significant as $T$ approaches $T_C$. Another parameter that
affects both $N(\tau)$ and $h(\tau)$ is the expansion rate of
the medium, {\it i.e.}, the
rate of change of temperature
between $\tau_c$ and $\tau$, which also depends on $\tau_c$. The overall
effect is that
both $N(\tau)$ and $h(\tau)$ rise
faster as compared to their homogeneous counterparts (see Fig. 8
for $\tau_c$=25 fm/c), particularly when $\tau_c$ is very large. (Compare
the left and right curves on Fig. 8.)
The increase in rates of
$N(\tau)$ and $h(\tau)$ is also larger for small $\sigma$ at large $\tau_c$.

For weak enough transition, the presence of inhomogeneity
may also affect several observables which can
be detected experimentally. For example,
the faster rate of hadronization at
large $\tau_c$ as compared to its homogeneous
counterpart will lower the amount of
entropy production, which, in turn, will affect the final hadron
multiplicity distributions. Although not studied here, the bubble size
distribution will also be affected by
the dynamics of nucleation \cite{ZABRO1}.
Since the nucleating bubble will act as a source of pion emission, the effect
of inhomogeneity can also be inferred through interferometry measurements.

In a cosmological context, the value of $\tau_c$ is much larger
than what was quoted here. Since the presence of inhomogeneities weakens the
transition, more critical bubbles will be nucleated per unit volume,
decreasing the inter-bubble distance,
$(d\approx N^{-1/3})$; the presence of subcritical bubbles can be thought as
seeds for nucleation. As a consequence, the transition will produce smaller
fluctuations in baryon number, protecting homogeneous nucleosynthesis.
Although the present study is indicative
enough of the reduction in the mean inter-bubble separation as compared
to homogeneous nucleation, a quantitative estimate would
require a more detailed analysis, including expansion. However, since the
cosmological expansion rate is typically much slower than the subcritical
bubble nucleation rate, we believe our results for the inter-bubble distance
will carry on in this case as well.

\section{Conclusions}
We have estimated the amount of phase mixing
due to subcritical hadron bubbles from very weak to
very strong first-order phase transitions.
With a reasonable set of values for the surface tension and
correlation length (as obtained from lattice QCD calculations), we found
that phase mixing is small at $T=T_C$ , building up as the
temperature drops further.
We have shown that
the system does not mix beyond the percolation threshold, allowing us
to describe the dynamics of the phase transition on the basis of
homogeneous nucleation theory with a reduced nucleation barrier.
Accordingly, we have found an enhancement in the nucleation
rate which further reduces the amount of supercooling. Although we have not
included cosmological expansion in our analysis, we believe that our results
indicate that the presence of an inhomogeneous background of subcritical
bubbles will decrease the inter-bubble mean distance, and thus the
fluctuations in baryon number which could damage homogeneous nucleosynthesis.

We have assumed that the equilibration time-scale for
subcritical fluctuations is much larger than the cooling time-scale
of the system. This may be the case for a quark-hadron phase transition
in the early universe, where the expansion rate is quite slow. In the
case of QGP produced at RHIC and LHC, the cooling rate is much faster
than cosmological time-scales, and the subcritical
bubbles density distribution may not attain full equilibration.
We are presently investigating this issue in more detail. However, the
present results should provide an
upper bound on the fraction of subcritical hadron bubbles and their
effect on supercooling and nucleation rates.

\acknowledgements

MG is partially supported by a National
Science Foundation Grant PHYS-9453431.

\newpage
{\bf\large Figure Captions}

Fig.~1 The effective potential as a function
of order parameter $\phi$ at, below and above $T_c$.

Fig.~2 Subcritical hadronic fraction $\gamma$ as a
   function of surface tension $\sigma$.

Fig.~3 The nucleation barrier $F_C/T$ for critical bubbles
with (solid line) and without subcritical bubble correction (dashed curve)
as function of temperature
for $\sigma=50$ MeV/fm$^2$ is shown in upper panel.
Corresponding subcritical hadron fraction $\gamma$
is shown in the lower panel.

Fig.~4 Same as Fig.~3 but  at $\sigma=30$ MeV/fm$^2$.

Fig.~5 Same as Fig.~3 but  at $\sigma=10$ MeV/fm$^2$.

Fig.~6  The temperature variation as a function of
proper time with (solid curve) and
without subcritical bubble correction (dashed curve)
for $\sigma$= 50 MeV/fm$^2$.

Fig.~7 Same as Fig.~6 but at  $\sigma$= 30 MeV/fm$^2$.

Fig.~8 (a) Density of nucleating bubbbles as a function of
proper time with (solid curve) and
without subcritical bubble correction (dashed curve)
for $\sigma$= 10 MeV/fm$^2$ (b) the hadronic fraction $\gamma$
as a function of $\tau$. The left curves are for $\tau_C=3.8$ fm/c and the right
curves for $\tau_C=25$ fm/c.

\newpage
\psfig{figure=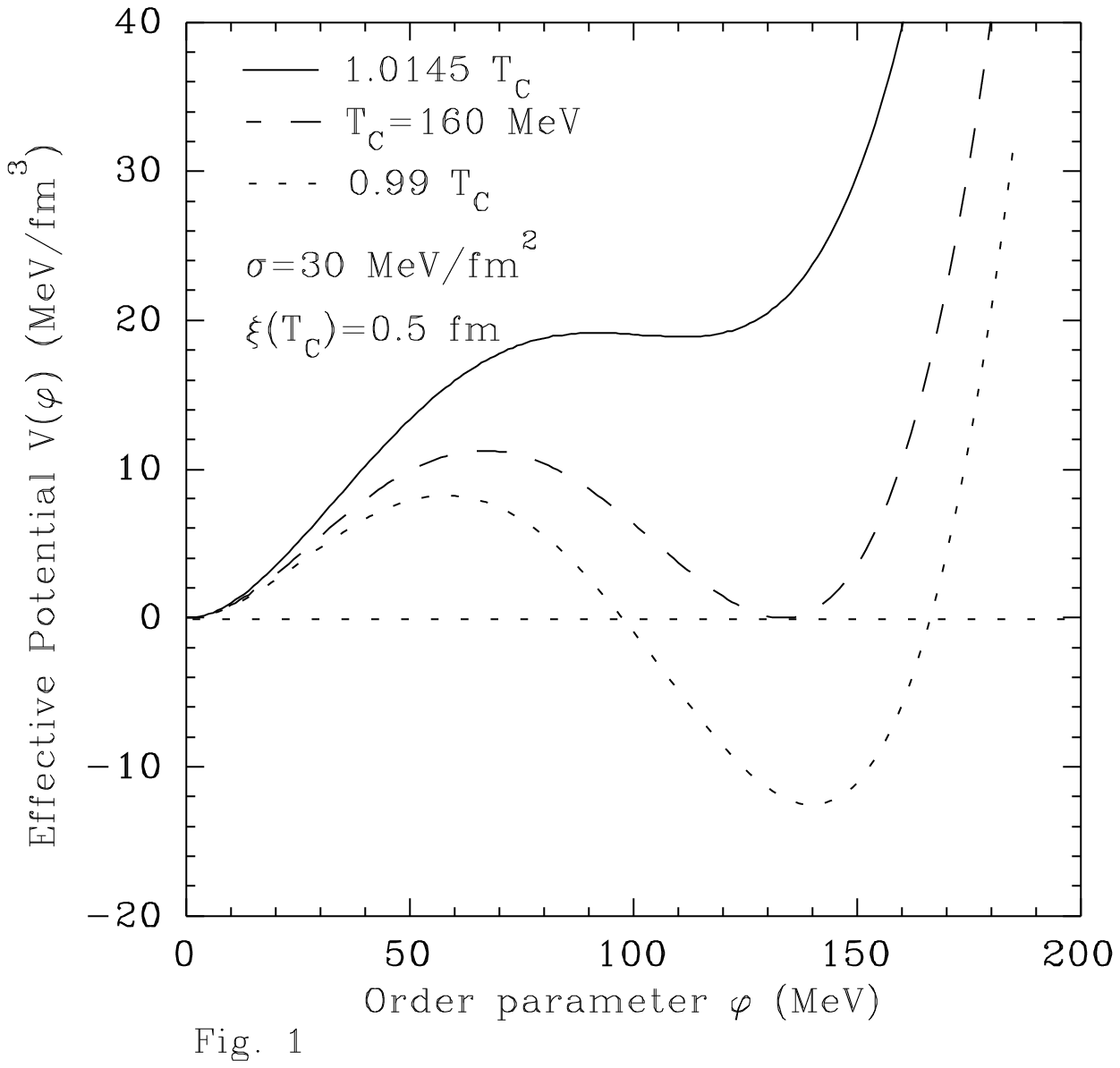}

\newpage
\psfig{figure=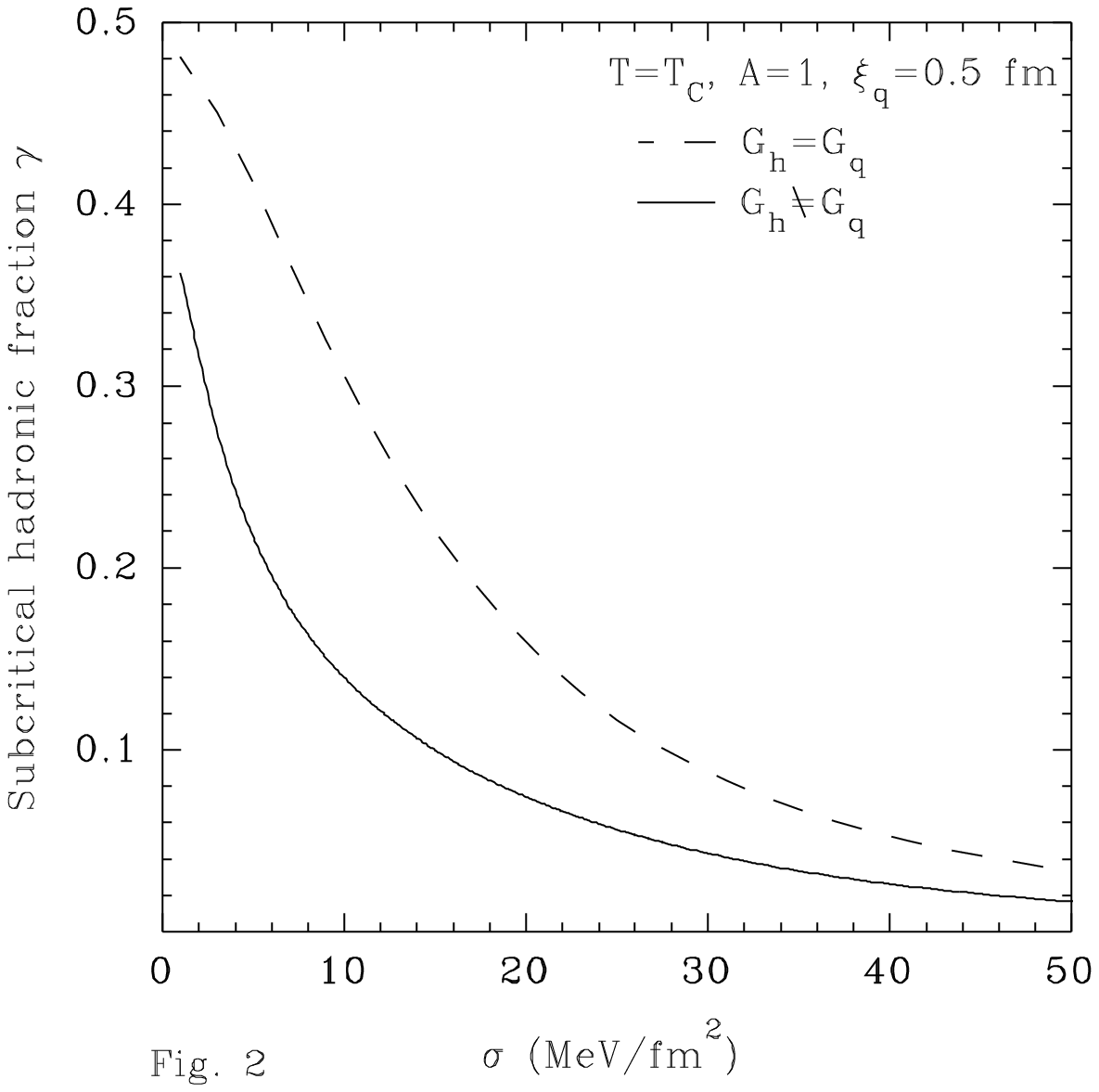}

\newpage
\psfig{figure=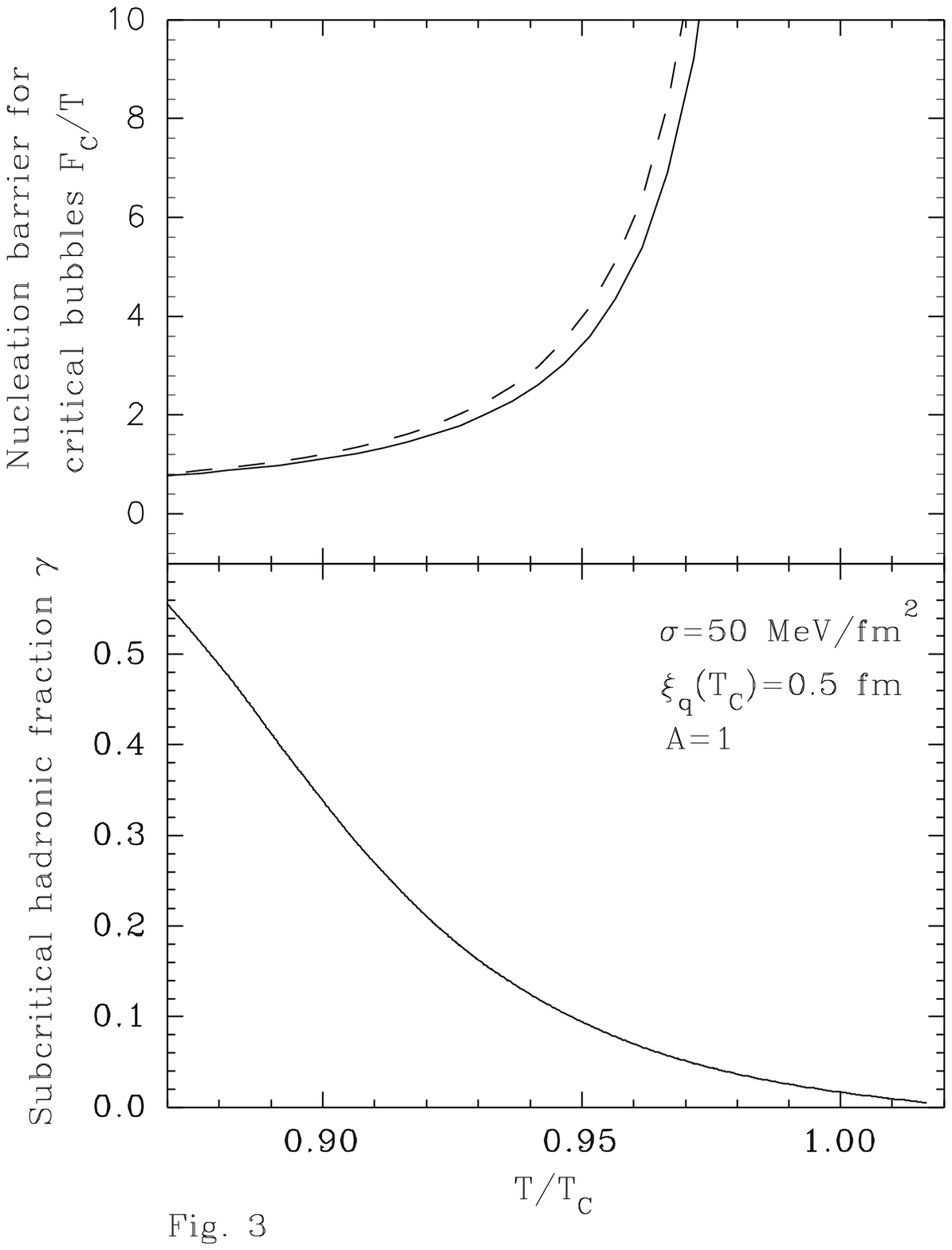}

\newpage
\psfig{figure=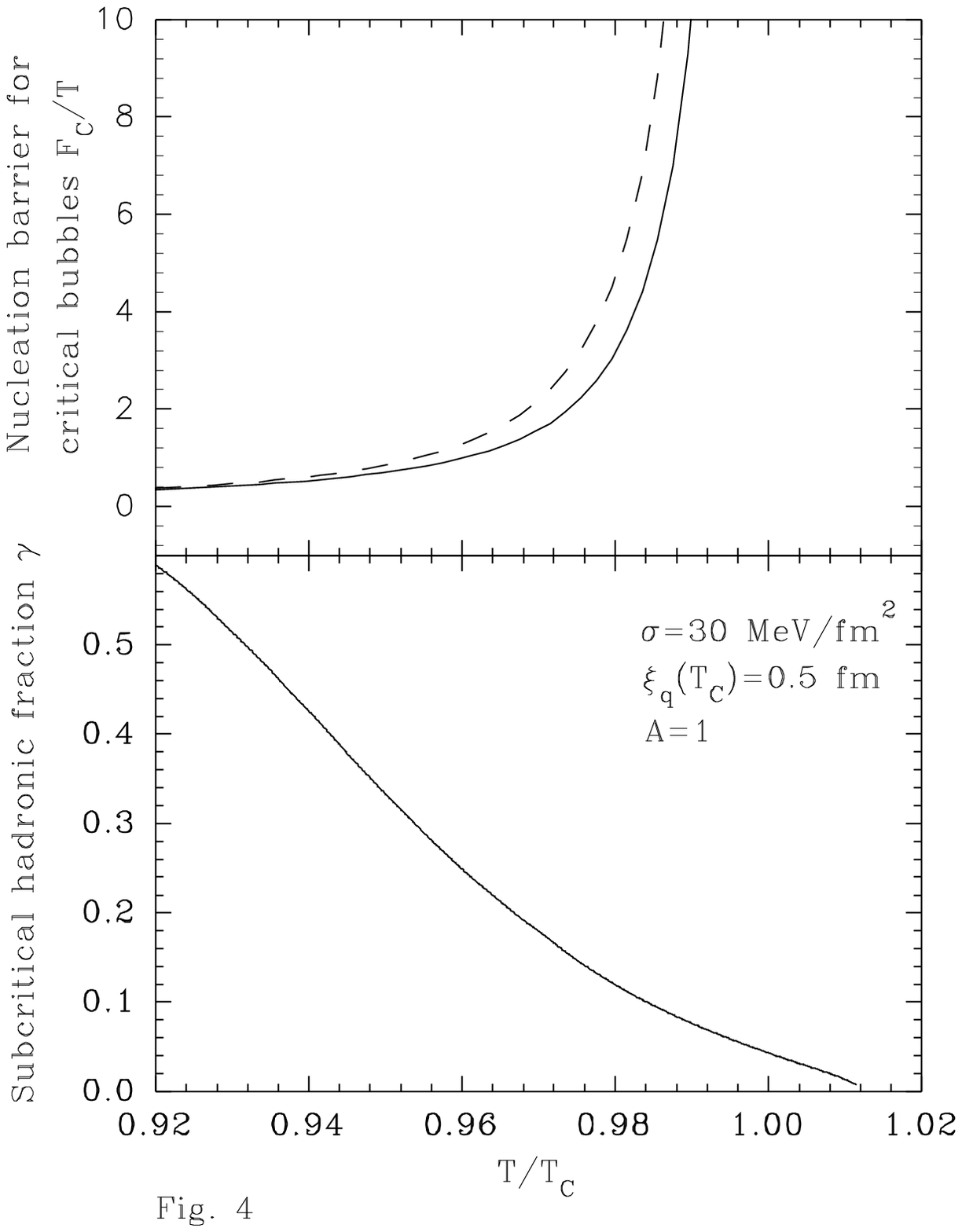}

\newpage
\psfig{figure=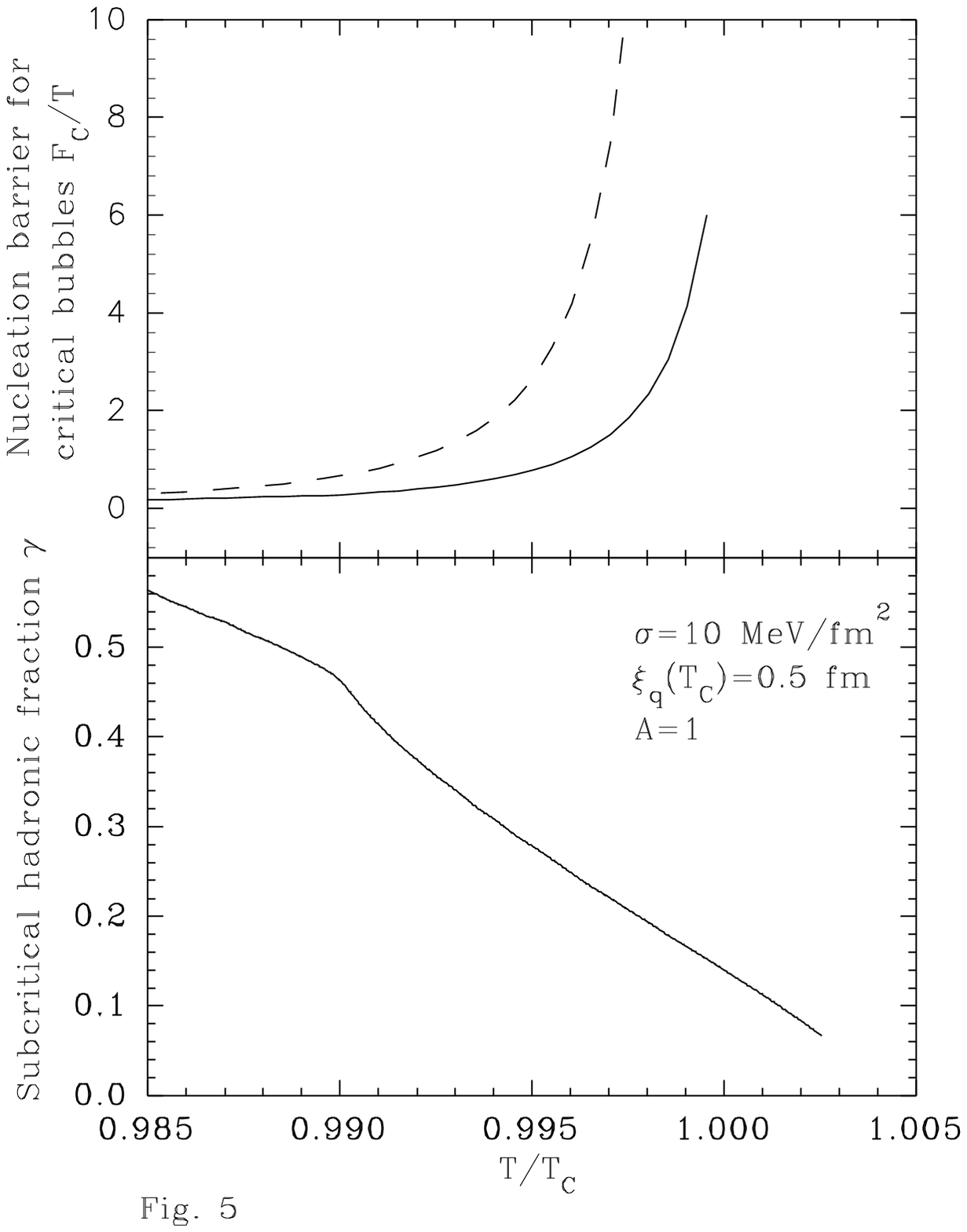}

\newpage
\psfig{figure=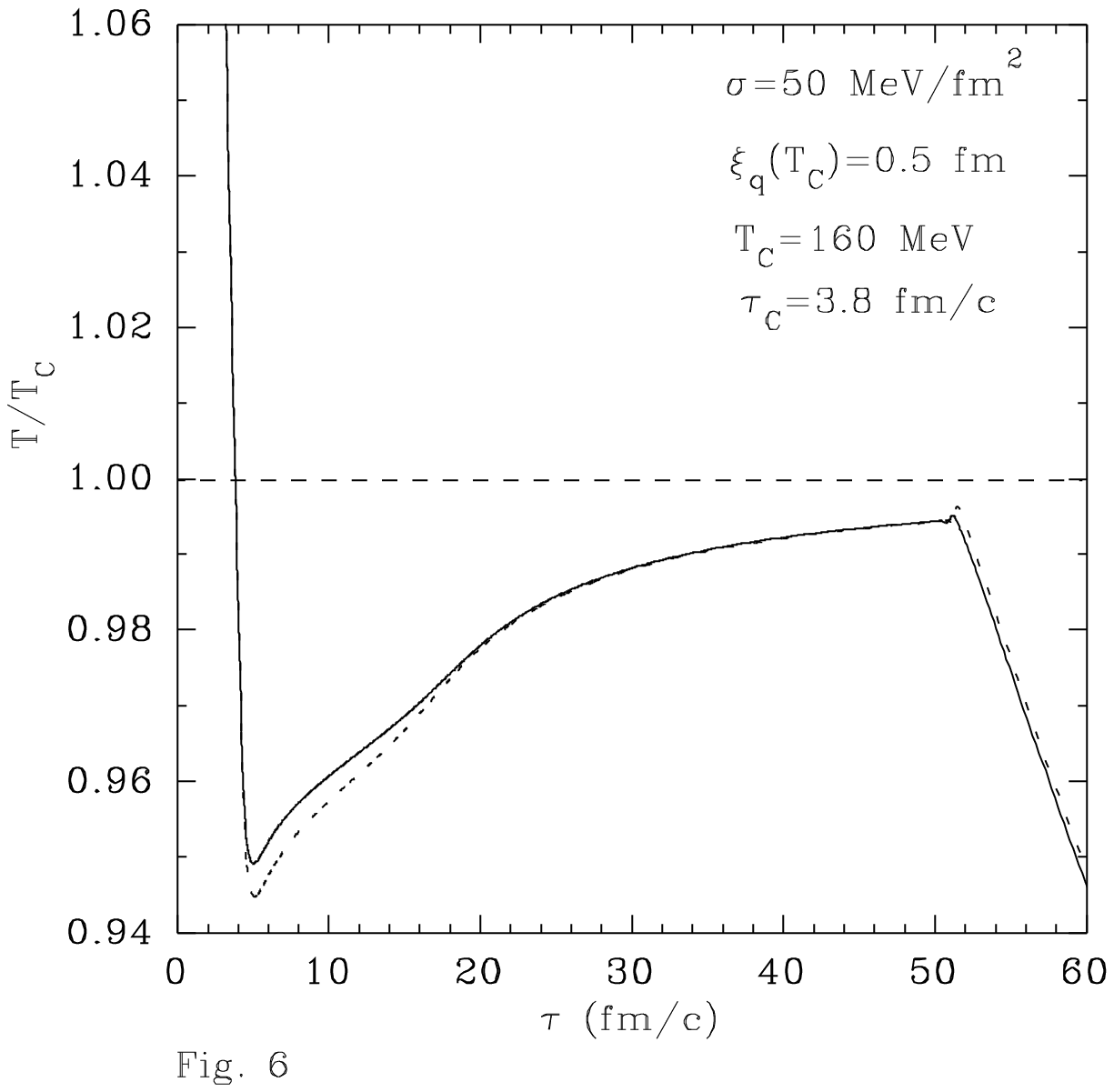}

\newpage
\psfig{figure=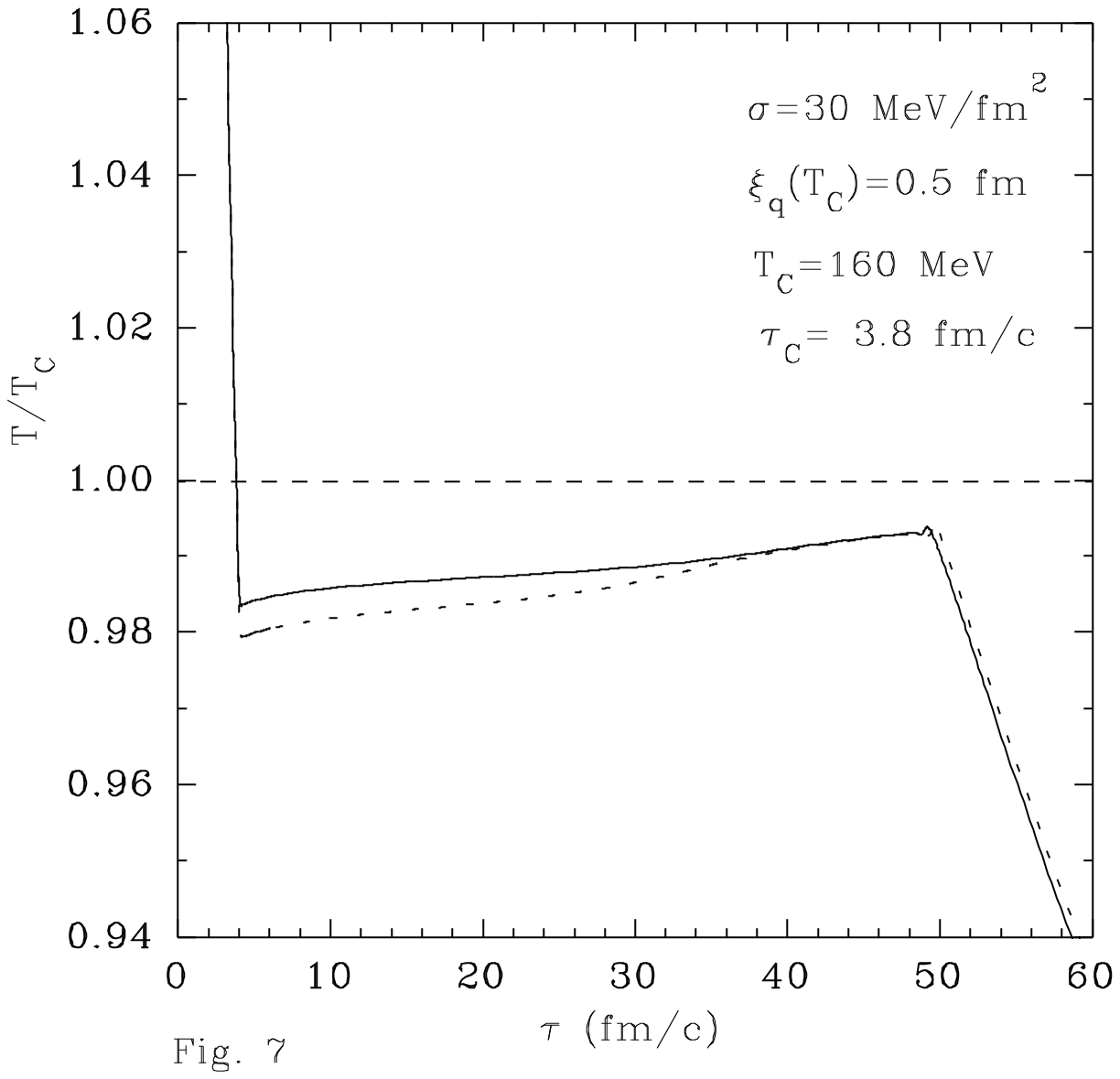}

\newpage
\psfig{figure=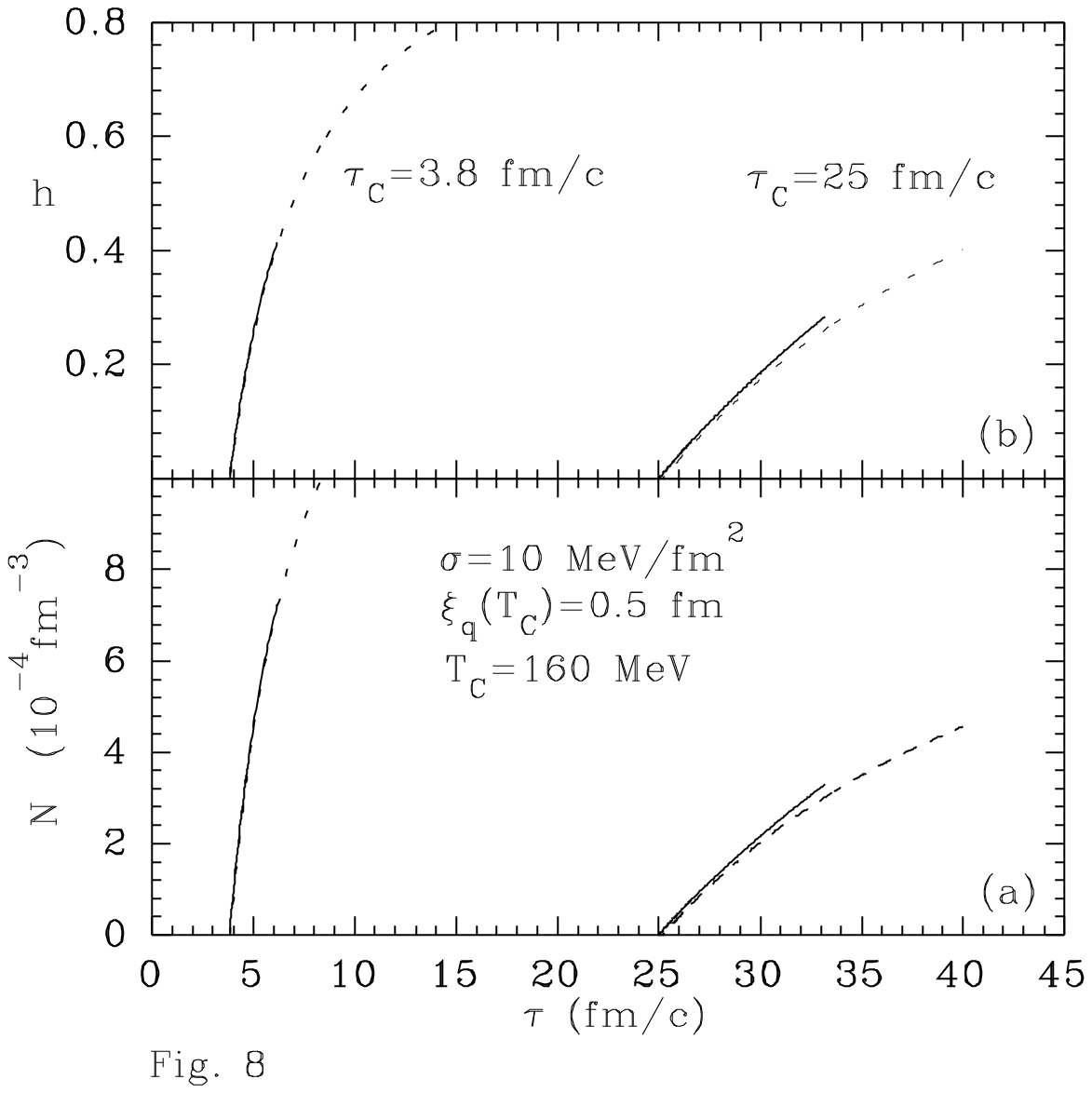}


\begin{thebibliography}{99}

\bibitem{LQCD1} Y. Iwasaki, K. Kanaya, S. Kaga, S. Sakai,
       and T. Yoshie in Lattice 94, Proceedings of international
       symposium, Germany, edited by F. Karsch et. al.
       [Nucl. Phys. B (Proc. Suppl.) {\bf 42}, 499 (1995)]

\bibitem{LQCD2} Y. Iwasaki, K. Kanaya, L. Karkkainen, K. Rummukainen,
        and T. Yoshie, Phys. Rev. D{\bf 49}, 3540 (1994);
        B. Beinlich, F. Karsch, and A. Piekert, Phys. Lett. B
        {\bf 390}, 268 (1997).

\bibitem{LANGER67}  J. S. Langer, Ann. Phys. (NY) {\bf 41}, 108 (1967);
{\it ibid.} {\bf 54}, 258 (1969).

\bibitem{LANGER73} J. S. Langer and L. A. Turski, Phys. Rev. A {\bf 8},
            3230 (1973).

\bibitem{CSER1} L. P. Csernai and J. I. Kapusta, Phys. Rev. D {\bf 46},
       1379 (1992).

\bibitem{CSER2} L. P. Csernai and J. I. Kapusta, Phys. Rev. Lett. {\bf 69},
               737 (1992).

\bibitem{ZABRO} E. E. Zabrodin, L. V. Bravina, L. P. Csernai, H. Stocker
and W. Griener,  Phys. Lett. B {\bf 423}, 373 (1998).

\bibitem{ZABRO1} E. E. Zabrodin, L. V. Bravina, H. Stocker
and W. Griener,  Phys. Rev. C {\bf 59}, 894 (1999).

\bibitem{SHUK} P. Shukla, S. K. Gupta, and A. K. Mohanty,
 Phys. Rev. C{\bf 59}, 914 (1999); hep-ph/9904345.

\bibitem{FULLER} G. M. Fuller, G. J. Mathews and C. R. Alcock,
      Phys. Rev. D {\bf 37}, 1380 (1988).

\bibitem{MEYER} B. S. Meyer, C. R. Alcock, and G. J. Mathews
      Phys. Rev. D {\bf 43}, 1079 (1991).

\bibitem{IGNAT1} J. Ignatius, K. Kajantie, H. Kurki-Suonio and
     M. Laine, Phys. Rev. D {\bf 50} 3738 (1994).

\bibitem{COND} T. Stinton III and J. Lister, Phys. Rev. Lett. {\bf 25},
            503 (1970).; H. Zink and W. H. de Jeu, Mol. Cryst. Liq. Cryst.
           {\bf 124}, 287 (1985).

\bibitem{GK} M. Gleiser and E. W. Kolb, Phys. Rev. Lett. {\bf 69},
       1304 (1992); M. Gleiser and M. Trodden, hep-ph/9911380.

\bibitem{PERCO} D. Stauffer and A. Aharony, {\it Introduction to
    Percolation Theory}, Taylor \& Francis, Washington, DC, 1992.

\bibitem{GH} M. Gleiser and A. F. Heckler,
        Phys. Rev. Lett {\bf 76}, 180 (1996).

\bibitem{AGAR} B. K. Agarwal and S. Digal, Phys. Rev. C{\bf 60}, 074007 (1999).

\bibitem{GKW} M. Gleiser, E. W. Kolb and R. Watkins,
        Nucl. Phys. B {\bf 364}, 411 (1991).

\bibitem{GELMINI} G. Gelmini and M. Gleiser, Nucl. Phys. B {\bf 419},
                   129 (1994).

\bibitem{GHK} M. Gleiser, A. F. Heckler, and E. W. Kolb, Phys. Lett. B
             {\bf 405}, 121 ((1997).


\bibitem{IGNAT2} J. Ignatius, K. Kajantie, H. Kurki-Suonio and
     M. Laine, Phys. Rev. D {\bf 49} 3854 (1994).

\bibitem{LINDE} A. Linde, Nucl. Phys. B {\bf 216}, 421 (1983).



\bibitem{RUGGERI} Franco Ruggeri and William A. Friedman,
         Phys. Rev. D {\bf 53}, 6543 (1996).

\bibitem{DYNA} P. Shukla, A. K. Mohanty, S. K. Gupta (hep-ph/0005219).

\bibitem{WEIN} S. Weinberg, Astrophys. J. {\bf 168,} 175 (1971).

\bibitem{MILLER} J. C. Miller and O. Pantano, Phys. Rev. D {\bf 40},
                 1789 (1989); {\bf 42}, 3334 (1990).

\bibitem{SHU} E. Shuryak, Phys. Rev. Lett., {\bf 68}, 3270 (1992).

\end{thebibliography}
\end{document}